\documentclass{acm_proc_article-sp}

\begin{document}

\title{ELECTRONIC HEALTH RECORDS AND CLOUD BASED GENERIC MEDICAL EQUIPMENT INTERFACE}

\numberofauthors{1} 
\author{
%
%
\alignauthor
Siddharth Srivastava, Ramji Gupta, Astha Rai, A S Cheema \\
\affaddr{Center for Development of Advanced Computing, Noida}
\email{\{siddharthsrivastava, ramjigupta, astharai, ascheema\}@cdac.in}
}

\maketitle
\begin{abstract}
Now-a-days Health Care industry is well equipped with Medical Equipments to provide accurate and timely reports of investigation and examination results. Medical Equipments available in market are made for specific tests suited for a particular laboratory leading to a wide variety of devices. The result viewing experience on console of these devices is not only cumborsome for medical staff but inefficient. Therefore, Medical Equipment Interfaces act as backbone of any Hospital Management Information System assisting in better management and delivery of test results. It also acts as a mode to collect data for further research and analysis. These equipments communicate via a fixed data format but compatibility among these formats is a major issue being faced in modern and legacy medical equipments. In this paper, we present a case study of designing and implementing a cloud based Generic Medical Equipment Interface(GMEI) along with the state of the art in such systems. This solution removes the burden of reentry of patient details into the Electronic Health Record(EHR) and thrives for accelerating EMR initiative in the country.
\end{abstract}


\section{Introduction}
Healthcare Industry has seen massive digitization and standardization in the past 5 years \cite{anderson2011digitization}. With the growing amount of patient related information, its efficient storage, proper management, timely access and security have now become imperative features for any healthcare solution. A substantial amount of patient information is in the form of test results such as biochemistry, pathology, and haematology. Another area of key importance is formation of Electronic Health Records \cite{slaveykov2013electronic, ambinder2005electronic}. Health Standards \cite{ferranti2006clinical} such as HL7 \cite{dolin2006hl7, beeler1998hl7}, Continuity of Care Document \cite{tekimplementation} etc. The Electronic Health Records have already been established as effective in many areas \cite{desroches2008electronic, hsiao2010electronic, linder2007electronic}. In this work, we develop a Cloud based Generic Medical Equipment Interfacing (GMEI) software solution addressing the problems stated above where Medical instrument interface is an automatic connection between analyzer and Host Computer for the quick exchange of information.

The GMEI at the back end, is supported by a sophisticated Health Information System (HIS). The Health Information System spans across all the major activities in a hospital such as Enquiry, Registration, Payment, OPD, Lab report etc. GMEI specifically deals with Lab Report which forms the largest chunk of information in an EHR. The other related information are utilized for maintaining the EHR of a patient. 
Contemporary studies\cite{nyasulu2014narrative, croft2011role, stephan2010using} have demonstrated the usefulness and importance of Medical Equipment Interfaces or the "Laboratory Information Systems" to a large extent. Therefore,  the main problems that have led to the need of the current work are:

\begin{itemize}
\item Lack of standardization in the data formats among medical equipments. 
\item In case of modern medical facilities, a few standard compatible (HL7, IHE etc.) machines are available. Yet, the availability is sparse and is not the same across various departments in hospitals. 
\item The test result data is stored over local storage attached with the machine in form of a Computer or dedicated storage. 
\item Due to local and heterogenous nature of data storage by different vendors, the unification of medical records of a patient for generation of EHR is difficult. This also presents a major drawback while updating and maintaining the EHR.
\end{itemize}

Recent research in Medical Equipment Interfaces or the "Laboratory Information Systems" has shown succesful migration of such systems to web interfaces and their need and importance \cite{cao2013opening, hong2011design, blaya2007web, runjie2005laboratory, xie2009design}. Moreover, in view of the growing need of analytics, context awareness, recommendations into Medical Equipment Interfaces \cite{braa1999interpretation, dube2003rigor, troshin2011pims}, there is a need to address the availability of data over cloud. Hence, the data not only from modern machines but legacy machines and equipments need to be made available for promoting and facilitating further research. 

\section{Methods}

\subsection{System Architecture}

\begin{figure*}
\centering
\includegraphics[width = 0.75\textwidth, height=0.75\textheight]{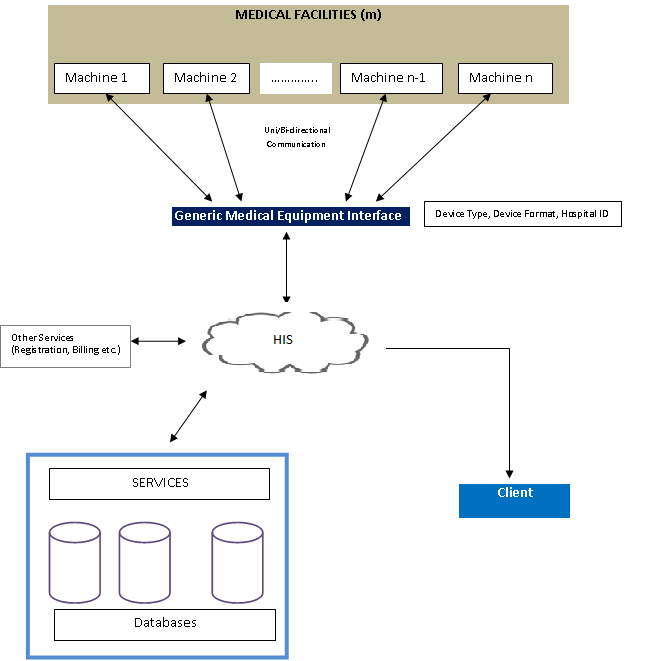}
\caption{System Architecture}
\label{fig:gmei}
\end{figure*}

Figure \ref{fig:gmei} depicts the general architecture of the developed system. GMEI can serve data from multiple machines(n) located at multiple(m) medical faciliities. The GMEI can be configured for data format specific to a device and hospital. The data received from the machines is organized into a prefixed global format by the GMEI. Subsequently, this is asynchronously sent to the cloud where the Health Information System maintains a uniform database for these records. 

This is in contrast to the traditional mode of communication where each LIS was vendor specific and hence the responsibility for data aggregation in a universally compatible format was left to the HIS. This type of flow resulted in additional complexities to the system. An LIS is aware of the incoming data format and related fields. We exploited this characteristics in GMEI and mapped each data format to a common database schema. This not only simplifies the overall architecture but also offloads task of machine specific parsing from the server. 

A major problem in laboratories is presence of legacy machines or machines with non-universal data formats. In our study, we found that over 95\% of the machines in various Government Hospitals across the country are not compliant to standards such as HL7, IHE etc. Most laboratories have heterogenous mixture of these formats i.e. a few machines are HL7 etc. compatible while others use different protocols. GMEI abstracts out this problem by assigning IDs to packet formats and devices. Another advantage of this approach is that if the device is upgraded or replaced, only the format ID at the GMEI needs to be reconfigured. The HIS can remain transparent to this modification. Since, an HIS can serve multiple hospitals, clinics etc. , the transparency further enahances the ease with which an EHR can be generated and shared among various entities. 

\subsection{Implementation of GMEI}

Following questions are needed to be addressed for planning implementation of GMEI:

\begin{itemize}
\item What would be the Mode of communicaton for transferring data and message between equipment and GMEI.
\item Will LIS generate work lists
\item Will samples be bar coded
\item How will LIS	match up results with sample ID
\item Are there any additional calculations that the end user may need for deductions from results
\item Can these calculations be automated in the interface
\item Does the technician need to review the results after analysis from machine
\end{itemize}

\subsubsection{Basic Requirement for Interfacing}
Analyzers capable of being interfaced to an LIS are found in all areas of the clinical laboratory: Bio-chemistry, Hematology, Urinalysis, Toxicology, Immunoassay, Coagulation, Microbiology, Blood Bank, and in another Labs. Nearly all of the routine clinical instruments are equipped with a serial RS-232 I/O port. Many new communication points like USB, TCP/IP port also provided. Some basic requirements neeedd to interface with instruments are enumerated below:

\textbf{\textit{A. Physical Hardware}} \\
The instrument must be equipped with an active input/output port to which computer devices may be connected. 

\begin{enumerate}
\item The host system must have a corresponding I/O port available.
\item On clinical analyzers, this is a serial synchronous RS-232 port. 
\item A connection cable is required, from the analyzer to an interface device or the host system connection point
\item 	Devices including bar code scanners, printers, environmental monitoring devices, etc.
\end{enumerate}

\textbf{\textit{B. Software}} \\
The host system software is operating system software and Generic Medical Equipment Interfacing application software. Operating system software is typically provided by a hardware manufacturer or a third party. Operating system software controls basic machine functions such as interaction with I/O devices, memory management, disk access, and creating the application software environment. GMEI software is the major product of the LIS software.

\textbf{\textit{C. Data Formats}} \\
There are many data formats, but for designing the GMEI, the data formats are categorized broadly into a few categories, and each data format from a machine is subclassified as a part. 

\textit{(i) RS-232 Serial Communication: }
These are the standards for communication with peripheral devices such as clinical analyzers. The parameters that classify this category are: 

\begin{itemize}
\item Serial communication parameters (baud rate, parity, data bits, stop bits, ASCII standard character set)
\item The "handshaking" protocol between host and a peripheral device for data exchange to occur: ACK/NAK
\item Message blocking: how does each side of the communication define where a message begins and ends. For example STX, ETX for start bit and end bit
\item Message structure (what will be the order of data fields, or where is the sample ID relative to the rest of the data record?)
\item Message content (what is the sample ID and what are the test names and results associated with this ID?
\end{itemize}

\textit{(ii) ASTM Protocol: }
Standards for communication between laboratory analyzers and information systems have been defined by ASTM (American Society for Testing and Materials). It aims towards simplification of connection between clinical analyzers and information systems. There are two standards that currently apply: E 1381-91 which defines protocol and E 1394-91 which defines format/content

\textit{(iii)HL7 standard: }
HL7 attempted to define a standard for communication between various parts of healthcare information systems (HIS), principally system-to-system communication. In the standard of their initial efforts, HL-7 became aware of parallel efforts at ASTM. While the HL-7 group is organized separately, the communication standard they have adopted is largely the work of another ASTM subcommittee within the E31 committee group

\subsection{GMEI: Modes of Operation}
GMEi can operate in two modes which are described in the following sections. 

\subsubsection{Unidirectional Mode}
Unidirectional mode is the most basic type of analyzer interface which reprsents the form of communication where the analyzer sends the data to the host. But, a unidirectional interface does not mean there is no communication at all coming down from the host or interface application to the analyzer. Most instruments require some form of handshaking or error checking to obtain data records. Such activities also involve sending ACK bit when collecting one result completed. 

Instrument performs its test and transmits results to the host interface system (GMEI). Many analyzer have been limited by design to unidirectional interfaces mode only. These analyzers are generally used for single test, batch or profile testing .

The result data transmitted from the analyzer generally includes the sample ID, the test names/ test codes, results, unit and normal range. Additional fields may include a variety of flag condition, sample type (blood, urine, STAT, control material), error check characters. Uni-directional interfaces are the easiest to implement.

\subsubsection{Bidirectional applications}
\begin{figure*}
\centering
\includegraphics[width=0.75\textwidth]{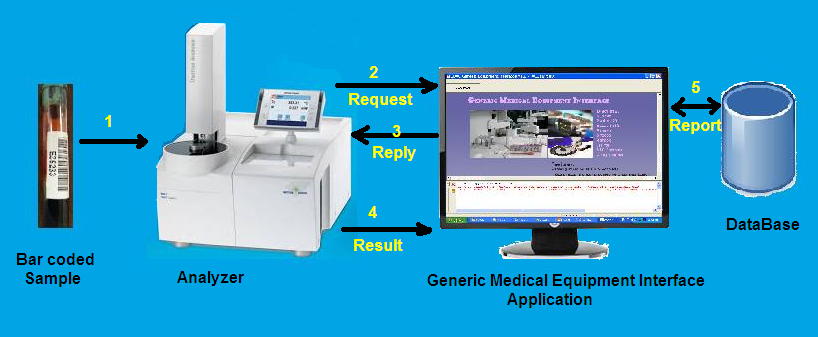}
\caption{Flow of Bidirectional Mode}
\label{fig:flow}
\end{figure*}

A bi-directional interface involves two-way communication between the analyzer and interface/host application. The host requests Sample ID and test order information from the HIS and sends the sample ID and test order information to the analzer. The flow  of bi-directional mode is shown in Figure \ref{fig:flow}. Support of bi-directional interface capability is mostly found in random access testing instruments, which can perform a different array of tests on each successive sample. 

Practically, to support a bidirectional communication, a DB9-to-DB9 rs232 cable with the specified pin out is required. Also,  a few flags in the communication settings for instance, baud rate, data size, parity, stop bit, time out, bcc check etc. need to be checked. In addition to flag based distinguishing system for unidirectional and bidirectional modes, an analyzer may also have CLASS A, CLASS B, ABX Format, ASTM Format etc. as choice for communication. These formats may be changed according to machine.  For seamless functioning of GMEI, the GMEI and Analyzer must synchronize according to data format otherwise bi-direction communication is not possible.

A bi-directional interface application saves the Lab technician time to program test orders into the analyzer and eliminate manual entry errors. This can result in a considerable improvement in analyzer productivity in a laboratory with very high sample load. Newer random access testing, bidirectionally interfaced analyzers will also incorporate bar code sample label scanning which provides  sample ID automatically. This can eliminate manual entry of sample IDs or coding of sample by tray and cup positions.

\subsubsection{GMEI: Workflow}

\begin{figure*}
\centering
\includegraphics{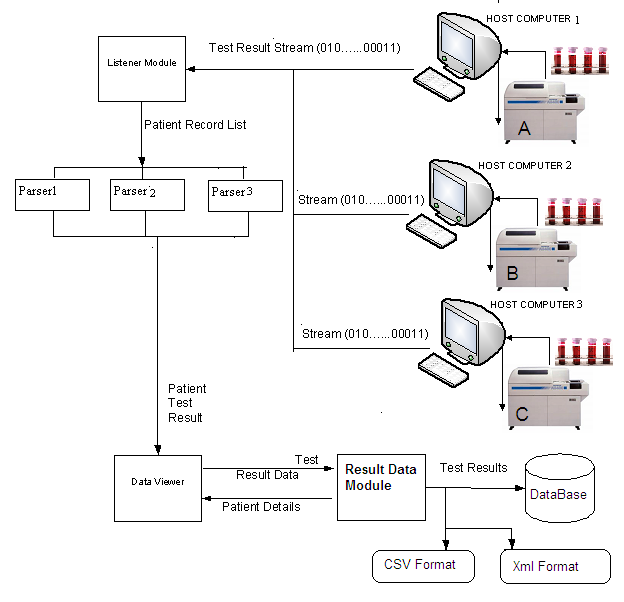}
\caption{Workflow of Generic Medical Equipment Interface}
\label{fig:workflow}
\end{figure*}

The workflow is shown in Figure \ref{fig:workflow} and is described below.

Generic equipment interface software is configured for different medical equipments (for example Machine A, Machine B and Machine C). To configure a medical equipment with GMEI, following entries are maintained in the database:

\begin{itemize}
\item Medical Equipment Master : to capture the machine information like communication protocol, data format, comport setting etc 
\item Machine-wise Test Parameter Master: used to capture the machine-wise test
\item User Master: for creating a of user 
\end{itemize}

The \textit{listener module} of application listens for the data from the medical equipment. Once the data is received, it is forwarded to appropriate parsing module according to equipment configuration settings. The packets are parsed asynchronously and submitted to the cloud for further processing.

It is important to note that some analyser machines do not have serial port interfacing and therefore provide test result output data into a log file format like .upl, .txt and .txt. GMEI handles this case by providing a file upload option as an input for bulk insertion from such devices, making it a generic interface in the true sense.

\section{Results}
 
The GMEI has been implemented in over 20 laboratories across the country which is working in sync with the HIS for delivering an end-to-end solution in various Government Hospitals. The major instruments in various laboratories which have benefited from GMEI are:
\begin{itemize}
\item Microbiology
\item Immunoassay
\item Coagulation
\item Hematology
\end{itemize}

In addition to this, the GMEI has been configured for the following medical equipments:

\begin{itemize}
\item OLYMPUS: AU 2700(Bi),AU 400(Bi), AU600
\item ROCHE: Elecsys2010,Cobas6000(Bi),Hitachi Modular(Bi)
\item BIOMERIEUX: Pentra120,Pentra60
\item SIEMENS: Dimension RxL(Bi), Advia Centaur 2120i
\item RANDOX: Daytona
\item BECKMAN COULTER: LH-500,LH-750,LH-780
\item Biorad D-10, Varient-II
\item STA Compact
\item Hematology Equipments: Hmx, ACL Top500, Sysmex XP-100
\item Blood Bank Equipments: Qwalys (*.upl File upload)
\end{itemize}

One of the busiest laboratories implementing GMEI has been using AU 2700 in bidirectional mode. Before incorporating GMEI to the workflow, the average load per week was 3000-4000 samples per week with an average time to complete the workflow was approximately 50 hours a week. Once the GMEI was fully incorporated in the workflow and integrated with HIS, the average time reduced to approximately 35 hours a week pertaining majorly to the reduction in sample processing, verification and report generation tasks. 

In many laboratories, more than one legacy machines were present. Moreover, since each machine belonged to different vendor, different local LIS was attached to each of them. This not only presented difficulty in training the staff and maintenance but also increased the overall cost of operation. With the introduction of GMEI, all the machines were able to migrate to a common interface and hence the training, maintenance and upgradation costs reduced drastically. 
	
\section{Conclusion}
During the development and implementation of the cloud based GMEI, its effectiveness in real time scenario were established. A general architecture was proposed for functioning of such systems and their integration with a HIS for maintaining an EHR of patients. Such systems would be extremely helpful in streamlining the process of standardization across the country and fulfilling various health initiatives by the Government. 

\bibliographystyle{abbrv}
\bibliography{sample}


\end{document}